\documentstyle[preprint,prb,aps]{revtex}
\begin{document}
\draft
\title{Inelastic lifetimes of confined two-component electron systems
in semiconductor quantum wire and quantum well structures}
\author{Lian Zheng and S. Das Sarma} 
\address{Department of Physics,
University of Maryland, College Park, Maryland 20742-4111}
\date{\today}
\maketitle
\begin{abstract}
We calculate Coulomb scattering lifetimes of electrons 
in two-subband quantum wires  
and in double-layer quantum wells
by obtaining the quasiparticle self-energy 
within the framework of the random-phase approximation
for the dynamical dielectric function.
We show that, in contrast to a single-subband quantum wire,  
the scattering rate 
in a two-subband quantum wire
contains contributions from
both particle-hole excitations and plasmon excitations.
For double-layer quantum well structures,
we examine individual 
contributions to the scattering rate 
from quasiparticle as well as acoustic and 
optical plasmon excitations at different 
electron densities and layer separations. 
We find 
that the
acoustic plasmon contribution in the two-component electron system does
not introduce any qualitatively new correction to the
low energy inelastic lifetime, and, in particular,
does not
produce the linear energy dependence
of carrier scattering rate as observed
in the normal state of high-$T_c$ superconductors.
\end{abstract}
\narrowtext
\section{introduction}
\label{one}
A central quantity in the theory of interacting electron 
systems is the quasiparticle inelastic lifetime, 
which is the inverse of the 
scattering rate due to electron-electron interaction
and determines the many-body broadening of the electron
spectral function
due to Coulomb interaction. The quasiparticle inelastic lifetime 
is important in understanding many physical phenomena 
in electron systems, such as tunneling,\cite{tun1}
transport,\cite{tra1} localization,\cite{loc1} etc.
The concept of the inelastic lifetime is also important 
in electronic device operations because it controls
the electron energy relaxation rate.\cite{hot1}
Over the past several decades, confined electron systems
in semiconductor heterostructures have been extensively 
studied for both fundamental and technological reasons.
The electron systems in high mobility GaAs/AlGaAs heterostructures
are especially suitable for 
studying electron-electron interaction  effects because of 
reduced impurity scattering resulting from the modulation-doping 
technique. In this article, we present our 
calculations of Coulomb scattering 
lifetimes of
electrons confined respectively in a
two-subband quantum wire and in a double quantum well structure
within the framework of the random-phase approximation (RPA). 
Our work is, therefore, a generalization of the
existing work in the literature
(which is mostly for the one-component electron systems) 
to the corresponding two-component one (quantum wire)
and two (quantum well) dimensional
electron systems.
We find that the scattering rate 
of a two-subband quantum wire
is quite different from that of 
a one-subband quantum wire.  
For a one-subband quantum wire, the
scattering rate is 
identically zero at small wavevectors and  
only plasmon excitation contributes at large wavevectors.
For a two-subband quantum wire, on the other hand, 
both quasiparticle and plasmon excitations contribute
and the scattering rate is non-zero even at small wavevectors
because of the quasiparticle contribution.
For a double quantum well electron system, 
we compare individual contributions 
to the scattering rate from quasiparticle as well as 
from acoustic and optical plasmon excitations at different
electron densities and interwell separations, and discuss 
implications of our results for other electronic systems
with layered structures. In particular, we show that 
acoustic plasmon excitations in a layered two-dimensional
electron system do not provide the 
linear energy dependence of carrier scattering rate
which has been observed
in the normal state of high-$T_c$ superconductors.\cite{htc1} 

Technological progress has made it possible to fabricate \cite{gon1}
high quality quantum wires where only the lowest one or two
subbands are populated by electrons. This has stimulated an increasing
interest \cite{rev1} in quantum wire structures. 
Much in the same way as quantum well structures have generated 
intense activity in pure and applied research in two-dimensional
electron systems, quantum wire structures have created the potential 
for new device applications \cite{rev1,sak1,mil1} 
and the opportunity to carry out experimental study on 
one-dimensional Fermi
systems, where many theoretical predications \cite{lut1,ben1}
can be tested.  Because of the low dimensionality,  quantum wire 
systems have many unique characteristics.
One of them
is that momentum and energy conservations
severely suppress low energy scattering processes and result in 
long lifetimes for electrons in quantum wires.
The conservation laws are much less restrictive 
on electron-electron scattering processes 
in two-subband quantum wires than in one-subband quantum wires,
and, as a result, the inelastic scattering rates
in a single subband quantum wire and in a two-subband quantum wire 
are quite different.
In a one-subband quantum wire, 
quasiparticle excitation does not contribute.\cite{ben1} The  
scattering rate is identically zero until the electron 
momentum becomes larger than a threshold value, 
where plasmon excitation begins to contribute.\cite{ben1}
For a two-subband quantum wire, 
both quasiparticle and plasmon excitations contribute to
the scattering rate. The
inelastic scattering rate 
is non-zero even at small wavevectors
because of the quasiparticle excitations. 
The inelastic scattering rate in a two-subband quantum wire
is, therefore, qualitatively different from the corresponding
one component situation, and, in fact, 
bears some ingredient
reminiscent of a higher dimensional system.
Our main motivation in studying the two-component quantum wire 
systems is two-fold:
(1) exploring its qualitative difference with the purely 
one-subband situation; and (2) the fact that the narrowest existing
doped semiconductor quantum wires have two occupied subbands.\cite{hwa2}
For simplicity, 
we only consider two-subband quantum wires where
electron-electron interaction potentials 
are independent of the subband index.
Our results are therefore strictly valid
only in quantum wires where the two subband occupations 
are due to Zeeman splitting.  
Qualitatively and even semiquantitatively, 
however, the results should also be applicable to 
other types of two-component quantum wires,
including two coupled-wire structures.

Inelastic electron lifetime due to Coulomb interaction
in two-dimensional systems has been well studied 
\cite{2ds1,2dd1,2dd2} in the literature.
There is a resurgent interest in 
the quasiparticle properties of layered two-dimensional electron systems 
due to the discovery of high-$T_c$ superconductors
which have layered structures.
One of the many unexpected behaviors observed in the high-$T_c$
superconductors is a linear energy dependence of
the normal state carrier scattering rate.
This anomalous linear
energy dependence of
the scattering rate is not expected from a single-layer uniform
two-dimensional electron gas with Coulomb interaction, 
where electron scattering rate is well known \cite{2ds1,2dd1,2dd2}
to be $\tau_k^{-1}\sim\xi_k^2\ln(|\xi_k|)$.
The essential new ingredient in a layered two-dimensional system 
is the presence of 
new low energy collective modes, the 
acoustic plasmon excitations.\cite{sds1}
There are recent suggestions \cite{gra1} 
that this anomalous linear energy dependence of 
the carrier scattering rate in 
the normal state of high-$T_c$ superconductors may
be explained by inelastic scattering processes due to
acoustic plasmon excitations. 
If this is true, {\it i.e.} if acoustic
plasmon scattering gives rise to a linear
energy dependence of the inelastic Coulomb scattering
rate, then the theoretical implications are tremendous 
because it is generally assumed that the standard many-body
Fermi liquid theory is not capable of explaining this linear dependence
( without invoking very special non-generic properties, for example,
van Hove points or hot spots, of the Fermi surface).
To further investigate this important issue,
we calculate the inelastic lifetime of two-dimensional electrons 
in a double quantum well structure
using the leading order many-body perturbative GW approximation.  We compare 
the contributions to the scattering
rate from acoustic plasmon excitations with contributions
from quasiparticle 
and optical plasmon excitations at different 
electron densities and interwell separations.
We show that the scattering rate due to the acoustic plasmon excitation
is, in general, small, and
does not provide either a qualitative or a quantitative
explanation for the linear energy dependence of  
carrier lifetimes in the normal state of high-$T_c$ superconductors. 
We therefore conclude that, contrary to the claims of Ref. 17,
layered structure by itself is not capable of producing a linear energy
dependence in the Coulomb scattering rate of two dimensional 
electron systems.

In Sec. \ref{lifet}, we briefly sketch the procedure
we use for calculating
the inelastic lifetime and then present and discuss 
our calculated results of electron lifetimes in a two-subband quantum wire
and in double quantum wells. 
A short summary in Sec. \ref{sum} concludes this article.

\section{inelastic lifetimes of two-component electron systems}
\label{lifet}
\subsection{Formalism}
The inelastic lifetime is obtained from a calculation of
electron self-energy due to Coulomb interaction. 
In this work, we approximate the self-energy 
by the leading perturbative term in an expansion in the
dynamically screened RPA exchange
interaction.\cite{mah1}
This approximation, commonly referred to as the GW approximation,\cite{mah1}
is extensively employed in calculating electronic many-body 
effects and is exact in the high density limit.
We restrict ourselves to the case where Coulomb  
interaction does not change the electron subband (layer) index.
Under this condition, the electron self-energy 
is diagonal with respect to the subband (layer) indices,
{\it i.e.} we assume that there is no Coulomb interaction induced 
intersubband (interlayer) scattering. \cite{rkl}
In the Matsubara formalism, the electron self-energy within
this GW approximation we employ is given by
\begin{equation}
\Sigma_{ii}(k,ip)=-{1\over\nu}\sum_{\bf q}{1\over\beta}
\sum_{i\omega}V^{scr}_{ii}(q,i\omega){\cal G}^0_{ii}
({\bf q}+{\bf k},i\omega+ip),
\label{equ:e1}
\end{equation}
where $\nu$ is the volume of the electron gas, 
$\beta$ is the inverse of temperature,
$i$ is the subband (layer) index, ${\cal G}^0$ is Green's function
of non-interacting electrons, and $V_{ii}^{scr}$ is the dynamically screened
intrasubband (intralayer) electron-electron interaction potential 
for which an explicit expression will be given below.
Inelastic lifetime of an electron with momentum $k$ in subband (layer) $i$
is obtained from the retarded self-energy by $\tau_{ki}^{-1}=
-(2/\hbar){\rm Im}\Sigma_{ii}(k,\xi_{ki})$, where
$\xi_{ki}=\hbar^2k^2/(2m)-\epsilon_{Fi}$ is the bare energy of the electron 
measured with respect to
the Fermi energy of subband (layer) $i$. The imaginary  
part of the retarded self-energy at zero temperature 
is given by Eq.(\ref{equ:e1}) as \cite{mah1}
\begin{equation}
{\rm Im}\Sigma_{ii}(k,\xi_{ki})={1\over\nu}\sum_{\bf q}
{\rm Im}\left[ V_{ii}^{scr}(q,\xi_{{\bf q}+{\bf k}i}
-\xi_{ki})\right]
\left[\theta(\xi_{ki}-\xi_{{\bf q}+{\bf k}i})
-\theta(-\xi_{{\bf q}+{\bf k}i})\right],
\label{equ:e2}
\end{equation}
where $\theta$ is step function.
The screened electron-electron
interaction potential $V_{ii}^{scr}$ in the above expression
is given by \cite{lia1}
\begin{equation}
V_{11}^{scr}(q,\omega)={v_a(q)+\chi_{22}^0(q,\omega)
\left( v_a^2(q)-v_b^2(q)\right)
\over 1-v_a(q)\left(\chi_{11}^0
(q,\omega)+\chi_{22}^0(q,\omega)\right)
+\chi_{11}^0(q,\omega)\chi_{22}^0
(q,\omega)\left( v_a^2(q)-v_b^2(q)\right)},
\label{equ:v1}
\end{equation}
where $v_a$ and $v_b$ are unscreened intra- and intersubband (layer)
Coulomb interaction potential, 
and $\chi_{ii}^0$ is density-density response function of 
non-interacting electrons for which analytical expressions are known.
\cite{ben1,vin1} Equations (\ref{equ:e2}) and (\ref{equ:v1})
form a complete description for the electron self-energy,
from which the inelastic electron lifetime can be obtained. 
The self-energy in Eq.(\ref{equ:e2})
contains two kinds of contributions. One is the quasiparticle 
contribution which occurs when either Im$\chi_{11}^0$ or
Im$\chi_{22}^0$ are nonzero.  The other is the plasmon contribution
which occurs when both Im$\chi_{11}^0$ and Im$\chi_{22}^0$ are zero 
and the denominator of the expression in Eq.(\ref{equ:v1}) vanishes.
For a two-component electron gas, there are in general \cite {sds1} two
branches of plasmon mode.
In the following, we shall apply the above formalism to
two-component quantum wire and quantum well structures and
investigate the importance of individual scattering processes
due to quasiparticle and plasmon excitations in these
systems.

\subsection{Inelastic electron lifetime in a two-subband quantum wire}
In the following, we evaluate electron inelastic lifetime
due to Coulomb interaction
in a two-subband quantum wire.  Our purpose is to compare 
the result of the two-subband quantum wire with that of a 
one-subband quantum wire and illustrate the generic features
of the inelastic lifetime in the two-subband case.

We consider the case where the two subbands are occupied 
by different numbers of electrons, so each subband $i$
has its own Fermi wavevector 
$k_{Fi}$ and Fermi energy $\epsilon_{Fi}$ ($i=1,2$).
As discussed before, we restrict ourselves to the case where 
the Coulomb potential is subband independent,
{\it i.e.} 
$v_a(q)=v_b(q)$ in Eq.(\ref{equ:v1}).
This is not a particularly severe approximation in view of the 
fact that the exact matrix elements of Coulomb interaction in a quantum wire
depend on the details of subband confinement, which, in general, 
are not known for quantum wires.
In our calculation, the bare potential $v_a(q)$ 
is obtained by using the infinite-well confinement model.\cite{str1}
The input materials parameters in our calculation are for 
GaAs-based quantum wires.\cite{ben1,hwa2}

In Fig.\ref{f1}, the excitation spectrum of the two-subband 
quantum wire is shown,
where the plasmon modes are represented by solid-lines,
and the quasiparticle excitations
are confined within the shaded area.
The spectrum consists 
of two regions of quasiparticle excitations and two branches
of plasmon excitations. Both the plasmons are approximately linear 
modes in the long wavelength limit.
The high energy mode exists for all 
values of wavevectors, while the low energy mode 
exists only within a limited range of wavevectors.
The most characteristic feature of 
the spectrum 
is that low energy quasiparticle excitations are prohibited,
{\it i.e} the phase space within $0<q<2k_{Fi}$ 
and $0<\omega<{\hbar^2\over2m}(2qk_{Fi}-q^2)$ ($i=1,2$)
is excluded from the quasiparticle excitation spectrum.
Both quasiparticle and plasmon 
excitations contribute to the inelastic scattering rate 
in the two-subband quantum wires as we shall show below.

In Fig.\ref{f2}, the inelastic scattering rate of a two-subband  
quantum wire is shown 
as a function of electron momentum $k$. 
The distinct feature of a
one-subband quantum wire is 
the total suppression, due to restrictions from energy 
and momentum conservations, of the quasiparticle contribution 
to the inelastic scattering rate.\cite{ben1}
Consequently, the scattering rate of a single-subband quantum wire
is identically zero until the
plasmon contribution is turned on at $k>k_c>k_F$.
Occupation of the second subband by electrons in a quantum wire
opens up additional scattering channels and 
makes the energy-momentum conservation less restrictive
on scattering processes. 
As a result, 
both quasiparticle and plasmon excitations contribute to the
inelastic scattering rate.
We see, from Fig.\ref{f2}, that the inelastic scattering
rate has contributions from quasiparticle excitations
at small wavevectors, 
from the second (low-energy) plasmon excitation 
at larger wavevectors, 
and from the first (high energy) plasmon excitation
at even larger wavevectors.
Compared with that of a single-subband quantum wire,
the inelastic scattering rate 
in a two-subband quantum wire contains some
ingredient of the corresponding
two-dimensional system. 
For quantum wires with increasing widths,
one may expect an eventual 
crossover to higher-dimensional behavior
as a large number of subbands become occupied by electrons.

In Fig.\ref{f3}, we compare contributions to 
the inelastic scattering rate
from the quasiparticle, the first (high-energy)
plasmon, and the second (low-energy) plasmon excitations at different 
occupations of the second subband.
We see that the quasiparticle excitation contributes to the inelastic 
scattering rate for $k\leq k_{F2}$,
and the second plasmon mode contributes 
for $k\geq k_{F2}$ while
the first plasmon mode contributes for
$k>k_c>k_{F1}$.
The momentum thresholds for
plasmon contributions
increase as the density of electrons in the second subband
increases.

The inelastic scattering rates in a
two-subband quantum wire and in a single-subband quantum wire
share qualitatively similar behaviors in some respects.
For a single-subband quantum wire \cite{ben1} under the ideal condition
of zero temperature and no disorder, Im$\Sigma(k_F,\omega)\sim 
\omega[\ln(|\omega|)]^{1/2}$ for $\omega\rightarrow0$,
which comes entirely from the plasmon contributions. 
Following the same procedure, one can show that
the same functional dependence, 
Im$\Sigma_{11}(k_{F1},\omega)\sim
\omega[\ln(|\omega|)]^{1/2}$ for $\omega\rightarrow0$,
is valid for a two-subband quantum wire as well,
which again arrives from the 
contribution of the first (high energy) plasmon 
excitation.
We like to point out that 
this asymptotic dependence of Im$\Sigma$,
which could imply non-Fermi-liquid properties, 
may not apply in real situations where 
temperature is finite
and disorder scattering is inevitably present.
More detailed discussion in this respect can be found in Ref. 11.
We should also mention that intersubband scattering ( which
is neglected in our model) leads to an off-diagonal self-energy contribution
whose imaginary part ( being quantitatively 
small for our two-subband model) has
a different asymptotic $\omega$-dependence which eventually produces 
the two dimensional behavior in the limit of
the occupancy of many one dimensional subbands.

\subsection{Inelastic electron lifetime in double quantum wells}
\label{dqw}
In the following, we investigate inelastic
scattering lifetime of electrons in 
double quantum wells.
We compare individual contributions
to the scattering rate from quasiparticle and plasmon 
excitations at different electron densities and 
interwell separations. 
The emphasis is on understanding the acoustic plasmon contribution
to the inelastic Coulomb scattering lifetime
and its implications
for other electronic systems with layered structures, such as
high-$T_c$ superconductors.
In particular, we want to correct some erroneous statements \cite{gra1}
on this point in the literature.

We consider the case where the two quantum wells
have equal electron densities.
The unscreened intra- and interwell Coulomb 
potentials in Eq.(\ref{equ:v1})
are $v_a(q)=2\pi e^2/q$ and $v_b(q)=v_a(q)\exp(-qd)$, where
$d$ is the interwell separation.  The effect of finite well-thickness
is ignored, as it does not change any of the results qualitatively,
although incorporating 
the effect into our calculations is straightforward.\cite{2dd2}
We neglect interwell tunneling effect and interwell Coulomb
scattering ({\it i.e.} the off-diagonal element of the Coulomb 
interaction).

In Fig.\ref{f4}, we show the excitation spectrum of a double quantum
well system,
where plasmon modes are represented by solid-lines,
and quasiparticle excitations
are within the shaded area.
The quasiparticle
excitation of a double quantum well system 
remains the same as that of a single quantum well system.
The essential 
new feature in a double quantum well system
is that there are two branches of plasmon excitations. 
In the present case of equal electron densities in the two layers, 
the plasmon 
dispersions $\omega_{\pm}(q)$ are 
solutions to $1=[v_a(q)\pm v_b(q)]\chi^0(q,\omega_{\pm})$,
where $\chi^0(q,\omega)$ is density-density response function
of a single-layer non-interacting 
two-dimensional electron gas.\cite{vin1}
In the long wavelength limit,
$\omega_-(q)$ is a linear acoustic plasmon mode,\cite{sds1}
whose contribution to the scattering rate is the major concern 
of our calculation. 

In Fig.\ref{f5}, we show the calculated
inelastic scattering rate of electrons
in double quantum wells,
along with each individual contribution from 
quasiparticle as well as acoustic and optical plasmon excitations.
Quasiparticle excitations contribute essentially at all values of 
wavevectors,
and it is the only non-zero contribution at small 
wavevectors.
Acoustic and optical plasmon excitations begin to contribute
when the wavevector is larger than the respective thresholds
$k_{ac}$ and $k_{op}$, where $k_F<k_{ac}<k_{op}$.
The contributions from acoustic and optical plasmon excitations 
are peaked within narrow windows of wavevectors just above 
the respective thresholds,
and diminish as the wavevector is further increased.  
The threshold wavevectors,
which depend on the parameters (such as electron densities, layer 
separation, etc.,) 
are $k_{ac}\sim1.5k_F$ and $k_{op}\sim2k_F$
for the particular choice of 
sample parameters of Fig.\ref{f5}. 
It is clear that neither plasmon excitation
contributes to the electron scattering close to the Fermi 
surface as they both become operative 
only above some threshold wavevectors.
Therefore, the inelastic electron
lifetimes close to the Fermi surface 
are determined entirely by quasiparticle excitations.

In a single quantum well system, inelastic 
scattering rate of electrons close to the Fermi surface
has an energy dependence Im$\Sigma(k,\xi_k)\sim\xi_k^2\ln
(|\xi_k|)$ for $\xi_k\rightarrow0$, which 
comes from the contribution by quasiparticle excitations.
\cite{2ds1,2dd2}
In a double quantum well system,  the scattering rate of electrons close
to the Fermi surface is also determined 
by quasiparticle excitations.  
Following a similar procedure, the scattering rate in
double quantum wells is found to have the same 
energy dependence Im$\Sigma(k,\xi_k)\sim\xi_k^2\ln
(|\xi_k|)$ for $\xi_k\rightarrow0$.
The fact that 
the inelastic Coulomb scattering lifetime in 
both a double quantum well system
and a single quantum well system has
the same asymptotic energy dependence 
is very easy to understand physically.
Scattering processes for electrons close to the Fermi
surface involve only long wavelength excitations, 
where the double quantum well system can 
be viewed as having zero interwell separation
because the interwell separation $d$ is finite whereas the effective
excitation wavelength is infinite--- 
equivalent to a single-well two-dimensional electron system
with an additional degeneracy of $2$
(similar to the valley degeneracy in Si-MOS system.)

In Fig.\ref{f6}, we compare individual contributions
to the scattering rate from quasiparticle as well as acoustic 
and optical plasmon excitations at different electron densities. 
As the density decreases, the effective strength of interaction 
increases, 
all the contributions to the scattering rate 
increase, while their relative strengths do not change significantly.

In Fig.\ref{f7}, we compare the individual contributions to the scattering
rate from quasiparticle and plasmon excitations
at different interwell separations.
The influence of the well separation on the quasiparticle 
(plasmon) contribution
is relatively weak (strong).
The two plasmon excitations are determined by  
the effective interaction $v_\pm(q)$, respectively, 
where $v_\pm(q)=v_a(q)\pm v_b(q)$, depending strongly on the
well separation through the $e^{-qd}$ factor.
As the separation decreases, $v_+(q)$ increases,
so the optical plasmon contribution 
increases, and the threshold momentum $k_{op}$ also increases.  
The situation for the acoustic plasmon is the opposite.
As the well separation decreases, $v_-(q)$ decreases, 
so the acoustic plasmon contribution 
decreases, and the threshold $k_{ac}$ decreases as well.
It should be noted that the  
acoustic plasmon contribution diminishes quickly as the well separation
decreases, so the electron 
lifetimes for energies close 
to the Fermi surface are not affected by the acoustic plasmon excitations
at any well separations.
The acoustic plasmons contribute significantly to the inelastic 
scattering rate only at large layer separations and at higher energies.

We now briefly discuss implications of our results 
for high-$T_c$ superconductors. 
In fact, our results essentially have no implications 
for the observed linear energy dependence of carrier 
scattering rates in the normal state 
of high-$T_c$ superconductors except to correct 
some recent erroneous and misleading claims \cite{gra1} in the literature.
Many experiments 
suggest linear energy and temperature dependence of lifetimes
for carriers close to the Fermi energy
in the normal state of high-$T_c$ superconductors.
It is well known that the anomalous linear
energy and temperature dependence of
the lifetime is not expected from a
Fermi-liquid type many-body theory for a single-layer uniform
two-dimensional electron gas.
In fact, the standard Fermi-liquid result 
\cite{2ds1,2dd2} for the scattering rate of
a two-dimensional electron gas is an
$(\epsilon-\epsilon_F)^2\ln(\epsilon-\epsilon_F)$ behavior
for $\epsilon$ close to $\epsilon_F$, which has been extremely
well-verified experimentally.\cite{tun1}
From the discussion above, it is clear that the acoustic plasmon
contributions 
in a layered
two-dimensional electron gas do not
produce a linear $(\epsilon-\epsilon_F)$ behavior
for the scattering rate either.
It is well-established \cite{sha1}
that both two- and three-dimensional electron
systems, at least within the standard perturbative many-body
picture, have scattering rates for electrons close to the Fermi surface
which
are essentially quadratic in energy, {\it i.e.}
$(\epsilon-\epsilon_F)^2$ within logarithmic corrections.
We show that the same is also true for a layered 
two-dimensional electron system.

\section{summary}
\label{sum}
In this work, we have studied inelastic scattering rates
due to electron-electron Coulomb interaction for 
two-component electron systems confined in 
semiconductor quantum wire 
and quantum well structures.
We showed that the scattering rate of a two-subband quantum wire
is quite different from that of a one-subband quantum wire
because momentum and energy conservations are 
less restrictive on scattering processes 
in two-subband systems where additional scattering channels 
are available.
For double quantum well systems, we compared 
individual contributions to the scattering rate from
quasiparticle as well as acoustic and 
optical plasmon excitations. We emphasized,
in particular, that the 
acoustic plasmon contributions in a layered electron system do not 
produce the linear energy dependence
of carrier scattering rate observed
in the normal state of high-$T_c$ superconductors.

\acknowledgments
This work is supported by the U.S.-ONR and the
U.S.-ARO.

\begin{figure}
\caption{ 
The excitation spectrum of a two-subband quantum wire electron system.
The plasmon modes are represented by solid-lines. 
The quasiparticle excitations
are confined within the shaded area. 
The electron densities in the two subbands 
$n_1=2.0\times10^5{\rm cm}^{-1}$ and $n_2=0.4n_1$.
The wire widths $L_y=L_z=100\AA$.
}
\label{f1}
\end{figure}

\begin{figure}
\caption{ 
The inelastic scattering rate of electrons in the first subband of
a two-subband quantum wire. 
The electron densities in the two subbands 
$n_1=2.0\times10^5{\rm cm}^{-1}$ and $n_2=0.4n_1$.
The wire widths $L_y=L_z=100\AA$.
}
\label{f2}
\end{figure}

\begin{figure}
\caption{ 
The inelastic scattering rate of electrons in the first subband of
a two-subband quantum wire due to the first plasmon excitations (a),
the second plasmon excitations (b), and the quasiparticle excitations (c)
at different occupations of the second subband.
The electron density in the first subband
$n_1=2.0\times10^5{\rm cm}^{-1}$.
The wire widths $L_y=L_z=100\AA$.
}
\label{f3}
\end{figure}

\begin{figure}
\caption{ 
The excitation spectrum of a double quantum well electron system.
The plasmon modes are represented by solid-lines. The quasiparticle excitations
are confined within the shaded area.
The electron densities in dimensionless unit 
$r_{s1}=r_{s2}=2$. The interwell distance 
$d=2.3a_B$.  
(In GaAs-based materials, 
$a_B\sim100\AA$, $r_{s1}=r_{s2}=2$ corresponds to 
$n_1=n_2\sim8\times10^{10}{\rm cm}^{-2}$.)
}
\label{f4}
\end{figure}

\begin{figure}
\caption{ 
The inelastic scattering rate of electrons in a double quantum well system.
The electron densities $r_{s1}=r_{s2}=1.4$.
The interwell distance $d=1.5a_B$.
}
\label{f5}
\end{figure}

\begin{figure}
\caption{ 
The inelastic scattering rate of electrons in a double quantum well system
due to the optical plasmon excitation (a), the acoustic plasmon excitation (b),
and the quasiparticle excitation (c)
at different electron densities.
The solid-line is for $r_{s1}=r_{s2}=1$ and the dotted-line
is for $r_{s1}=r_{s2}=2$.
The interwell distance $d=1.1a_B$.
}
\label{f6}
\end{figure}

\begin{figure}
\caption{ 
The inelastic scattering rate of electrons in a double quantum well system
due to the optical plasmon excitation (a), 
the acoustic plasmon excitation (b),
and the quasiparticle excitation (c)
at different well separations.
The solid-line is for $d=1.1a_B$ and the dotted-line
is for $d=0.6a_B$.
The electron densities $r_{s1}=r_{s2}=1$.
}
\label{f7}
\end{figure}

\end{document}